\documentclass[twocolumn]{revtex4}
\usepackage{amssymb}
\usepackage{latexsym}
\usepackage{epsfig}

\usepackage{amsmath}
\begin{document}
\title{Effects of Anisotropy on the Sign-Changeable Interacting Tsallis Holographic Dark Energy}
\author{M. Abdollahi Zadeh$^{1}$\footnote{mazkph@gmail.com}, A.
Sheykhi$^{1,2}$\footnote{asheykhi@shirazu.ac.ir}, H.
Moradpour$^2$\footnote{h.moradpour@riaam.ac.ir}, Kazuharu Bamba$^3$\footnote{bamba@sss.fukushima-u.ac.jp}}
\address{$^1$ Physics Department and Biruni Observatory, College of Sciences, Shiraz University, Shiraz 71454, Iran\\
$^2$ Research Institute for Astronomy and Astrophysics of Maragha
(RIAAM), P.O. Box 55134-441, Maragha, Iran\\
$^3$ Division of Human Support System Faculty of Symbiotic Systems
Science Fukushima University, Fukushima 960-1296, Japan}

\begin{abstract}
A spatially homogeneous and anisotropic Bianchi type I universe is
considered while it is filled by pressureless dark matter (DM) and
Tsallis holographic dark energy (DE) interacting with each other
throughout a sign-changeable mutual interaction. Various infra-red
(IR) cutoffs are studied, and it has been obtained that while the
current universe can classically be stable for some cases, all
models display classical instability by themselves at the future
($z\rightarrow-1$). Moreover, we find out that some models can
cross the phantom line. In order to have a more comprehensive
study, the statefinder diagnostic and the
$\omega_D-{\omega}^{\prime}_{D}$ plane are also investigated
showing that the model parameters significantly affect the
evolution trajectories in the  $r-s$ and
$\omega_D-{\omega}^{\prime}_{D}$ planes.
\end{abstract}
\maketitle

\section{Introduction}
Cosmological observations such as type Ia supernova
\cite{Riess,Riess1}, WMAP \cite{Bennett,Bennett1} and Large Scale Structure
(LSS) \cite{Verde,Verde1,Verde2} indicate that our Universe is now experiencing
a phase of accelerated expansion \cite{roos}. This accelerated
phase of the Universe expansion is describable by a mysterious
source of energy (DE) whose energy density $\rho_{DE}$ and
pressure $P_{DE}$ filling about $0.73$ percent of the Universe,
while $\rho_{DE}+3P_{DE}< 0$ meaning that its equation of state
(EoS) parameter, $\omega_{DE}={P_{DE}}/{\rho_{DE}}$, must satisfy
the condition $\omega_{DE}<-{1}/{3}$
\cite{Bamba:2012cp,Nojiri:2010wj,Capozziello:2011et,Capozziello:2010zz,Bamba:2015uma,Cai:2015emx,Nojiri:2017ncd,
shey1,shey2,shey3,shey4,shey5,shey6,shey7,shey8,shey9,shey10}. An interesting attempt to find a physical origin for DE is
called the holographic dark energy (HDE) making a relation between
the system entropy and the UV cutoff (or equally the energy
density of quantum fields in vacuum) \cite{Cohen99}.

Since gravity is a long-range interaction, some physicists tried
to study the cosmological evolution by using the generalized
entropy formalism in which the Bekenstein entropy (as the backbone
of the HDE hypothesis \cite{Cohen99}) is not always met
\cite{non2,non15,non16,non17,non18,non19,non20,non21,non22,non23,
non13,non4,non5,non6,non7,non8,non9,non10,non11,non12,non14,eb,eb1,3,4,5,6,7,8,9,10,10a,11,11a,11b,cite1,cite2}.
Recently, using generalized entropy formalisms, some new HDE
models have been introduced \cite{THDE,smm,epjcr}. Tsallis
holographic dark energy (THDE) is one of these new attempts, based
on Tsallis entropy \cite{Tsallis}, which can provide suitable
description for the current universe in various cosmological
setups \cite{THDE1,THDE2,THDE3,THDE4,Sar}. It is also worth
mentioning that although THDE is not stable at the classical level
\cite{THDE}, a more comprehensive study on its stability may
consider the global features of the metric perturbations
\cite{Stabil1,Stabil2}.

On the other hand, observations admit a mutual interaction between
DE and DM \cite{Cai,Wei,Wei1,Kh,Kh1,Kh2,Kh3,Kh4,Kh5,Kh6,Abdollahi2016,Abdollahi2017,Abdollahi2017a},
and the $CMB+BAO+SN+H_{0}$ data suggests that the sign of mutual
interaction has probably been changed during the cosmic evolution
in the $0.45\leq z\leq 0.9$ interval \cite{Cai}. This result
motivates physicists to consider sign-changeable interactions,
including the deceleration parameter
\cite{Wei,Wei1,Kh,Kh1,Kh2,Kh3,Kh4,Kh5,Kh6,Abdollahi2016,Abdollahi2017,Abdollahi2017a}. Although the
FLRW metric is a powerful tool in modeling the universe, and is in
accordance with the cosmological principle \cite{roos}, it does
not compatible with the early universe anisotropy, and also the
anisotropy of the cosmos in scales smaller than $100$-Mpc and also
\cite{roos}. Thus, a comprehensive study on the cosmic evolution
should consider the anisotropy effects. The Bianchi models are
some interesting attempts to model the anisotropy of the early
universe \cite{khani1,khani2,khani3,khani4,khani5,khani6,khani7,khani8,khani9,khani10,khani11,khani12,khani13,khani14,khani15,khani16}. Motivated by the above arguments,
we are going to study the cosmic evolution of a Bianchi type I
(BI) model filled by DE and DM interacting with each other
throughout a sign-changeable interaction. In our models, THDE with
various IR cutoffs is used to model DE.

The paper is organized as follows. In the next section, we present
the general remarks on the BI universe filled by mutually
interacting DE and DM. The results of considering the Hubble
horizon as the IR cutoff is investigated in sections
~$\textmd{III}$. The cases using the event and particle horizons
as the IR cutoffs are studied in sections ~$\textmd{IV}$) and~
$\textmd{V}$, respectively. Two other famous IR cutoffs, including
the GO \cite{Li20041,Li20042} and Ricci \cite{Li2004,Li2004a,Gao} cutoffs,
will be addressed in sections.~$\textmd{VI}$ and~$\textmd{VII}$,
respectively. The last section is devoted to a brief summary.
\section{GENERAL FRAMEWORK}

The BI metric, which includes the anisotropy of early cosmos, is written as \cite{khani1,khani2,khani3,khani4,khani5,khani6,khani7,khani8,khani9,khani10,khani11,khani12,khani13,khani14,khani15,khani16}
\begin{equation}\label{metric}
ds^2=dt^{2}-A^{2}(t)dx^{2}-B^{2}(t)dy^{2}-C^{2}(t)dz^{2},
\end{equation}
where $A(t)$, $B(t)$ and $C(t)$ are functions of cosmic time.
Hence, the FLRW metric is recovered whenever $A=B=C$
\cite{khani1,khani2,khani3,khani4,khani5,khani6,khani7,khani8,khani9,khani10,khani11,khani12,khani13,khani14,khani15,khani16}. For this metric, $i$)
$V^3\equiv A B C$ denotes the spatial volume. $ii$)
$a=(ABC)^{\frac{1}{3}}$ is defined as the average scale factor.
$iii$) $H=\frac{1}{3}(H_1+H_2+H_3)$ is the generalized mean Hubble
parameter, where $H_1= \frac{\dot{A}}{A}$, $H_2=\frac{\dot{B}}{B}$
and $H_3=\frac{\dot{C}}{C}$ are called the directional Hubble
parameters in the directions of $x,y$ and $z$ axes, respectively
\cite{khani1,khani2,khani3,khani4,khani5,khani6,khani7,khani8,khani9,khani10,khani11,khani12,khani13,khani14,khani15,khani16}.

Now, consider a BI universe filled by a pressureless source (with
energy density $\rho_m$) and a DE candidate with
$T^{\mu}_{\nu}=diag[\rho_D,-\omega_D\rho_D,-\omega_D\rho_D,-\omega_D\rho_D]$
where $\rho_D$ and $\omega_D(\equiv\frac{p_D}{\rho_D})$ represent
the energy density and EoS parameter of dark energy, respectively,
and $p_D$ denotes the DE pressure. In this manner, the
corresponding Friedmann equations (the BI equations) take the form
\cite{Hossienkhani,Hossienkhani11,Hossienkhani12}
\begin{eqnarray}
3H^{2}-\sigma^{2}&=&\frac{1}{m_p^2}(\rho_{m}+\rho_D),
\label{ani1}\\
3H^2+2\dot{H}+\sigma^{2}&=&-\frac{1}{m_p^2}\left(p_D\right), \label{ani2}\\
\end{eqnarray}
in which $m_p^2=1/(8\pi G)$ and $\sigma$ are the Planck mass and
the shear scalar, respectively. Moreover,
$\sigma^2=1/2\sigma_{ij}\sigma^{ij}$ while
$\sigma_{ij}=u_{i,j}+\frac{1}{2}(u_{i;k}u^{k}u_{j}+u_{j;k}u^{k}u_{i})+\frac{1}{3}\theta(g_{ij}+u_{i}u_{j})$
is called the shear tensor, which describes the rate of distortion
of the matter flow, and $\theta=3H=u^{j}_{;j}$ denotes the scalar
expansion. Defining the critical density $\rho_{cr}=3m_p^2H^2$,
and introducing the dimensionless density parameters
\begin{eqnarray}\label{Omega}
\Omega_{m} &=&\frac{\rho_{m}}{\rho_{cr}}, \quad  \Omega_D
=\frac{\rho_D}{\rho_{cr}}, \quad \Omega_{\sigma}=
\frac{\sigma^2}{3H^2},
\end{eqnarray}
the first BI equation can be rewritten as
\begin{eqnarray}\label{Omega}
\Omega_m+\Omega_{\Lambda}=1-\Omega_{\sigma}.
\end{eqnarray}
It is worthwhile mentioning that the shear scalar is describable,
using the average Hubble parameter, as $\sigma^2=\sigma_0^2H^2$ in
which $\sigma_0^2$ is a constant \cite{khani1,khani2,khani3,khani4,khani5,khani6,khani7,khani8,khani9,khani10,khani11,khani12,khani13,khani14,khani15,khani16}. In the presence of a mutual interaction ($Q$) between
the cosmos sectors, the continuity equation is decomposed as
\cite{Wei,Wei1}
\begin{eqnarray}\label{conm}
&&\dot{\rho}_m+3H\rho_m=Q,\\
&&\dot{\rho}_D+3H(1+\omega_D)\rho_D=-Q\label{conD}.
\end{eqnarray}
Following \cite{Wei,Wei1}, we assume $Q=3b^2 q H \rho_D (1+u)$ where
$b^2$ is coupling constant, $u=\frac{\rho_m}{\rho_D}$ and $q$ is
the deceleration parameter defined as
\begin{equation}\label{deceleration}
q=-\frac{\ddot{a}}{a H^2}=-1-\frac{\dot{H}}{H^2}.
\end{equation}
Thus, whenever the universe phase expansion is changed, the
interaction sign is also changed. Indeed, for $Q<0$ ($Q>0$), there
is an energy flow from DM (DE) to DE (DM). At the classical level,
for a stable model, the sound speed square ${v}^{2}_{s}$ is
positive \cite{Peebles20031}. For DE candidate, which controls the
current universe dynamics, it is evaluated as
\begin{equation}\label{vs}
v_{s}^{2}=\frac{dP_D}{d\rho_D}=\frac{\dot{P}_D}{\dot{\rho}_D}=\dfrac{\rho_{D}}{\dot{\rho}_{D}}
\dot{\omega}_{D}+\omega_{D}.
\end{equation}
A new set of completely geometrical parameters $\{r,s\}$, dubbed
the statefinder, has been introduced by Sahni et al \cite{Sahni}
\begin{equation}\label{statefinder}
s=\frac{r-1}{3(q-1/2)},~~~~~~~~~~ r=\frac{\dddot{a}}{aH^3},
\end{equation}
which obviously show the statefinder pair depend only on the scale
factor and it\textquoteright s time derivatives up to the third
order. Combining the above relation with each other, we can also
write
\begin{equation}\label{rr2}
r=2q^2+q-\frac{\dot q}{H}.
\end{equation}
Another way to analysis the cosmic evolution has been introduced
in Ref.~\cite{Caldwell}, based on the set of parameters
$\{\omega_D,{\omega}^{\prime}_{D}\}$ (prime denotes derivative
respect to $x=lna$). This approach works in the
$\omega_D-{\omega}^{\prime}_{D}$ plane, in which
${\omega}^{\prime}_{D}>0$ and $\omega_D<0$ present the thawing
region, while ${\omega}^{\prime}_{D}<0$ and $\omega_D<0$ present
the freezing region \cite{Caldwell}.
\begin{figure}[htp]
\begin{center}
\includegraphics[width=8cm]{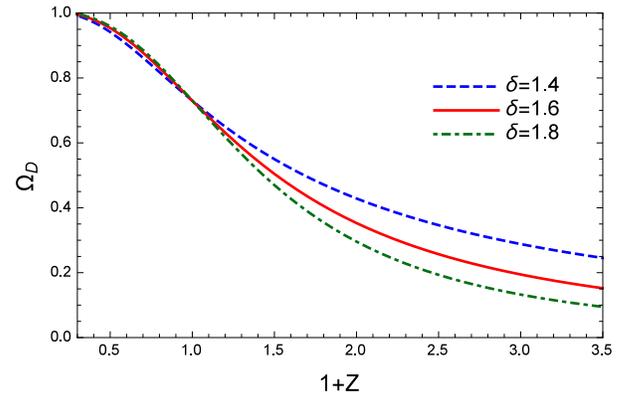}
\caption{$\Omega_D$ versus redshift parameter $z$ for
sign-changeable interacting THDE with Hbbble radius as the IR cutoff. 
}\label{Omega-z1}
\end{center}
\end{figure}

\begin{figure}[htp]
\begin{center}
\includegraphics[width=8cm]{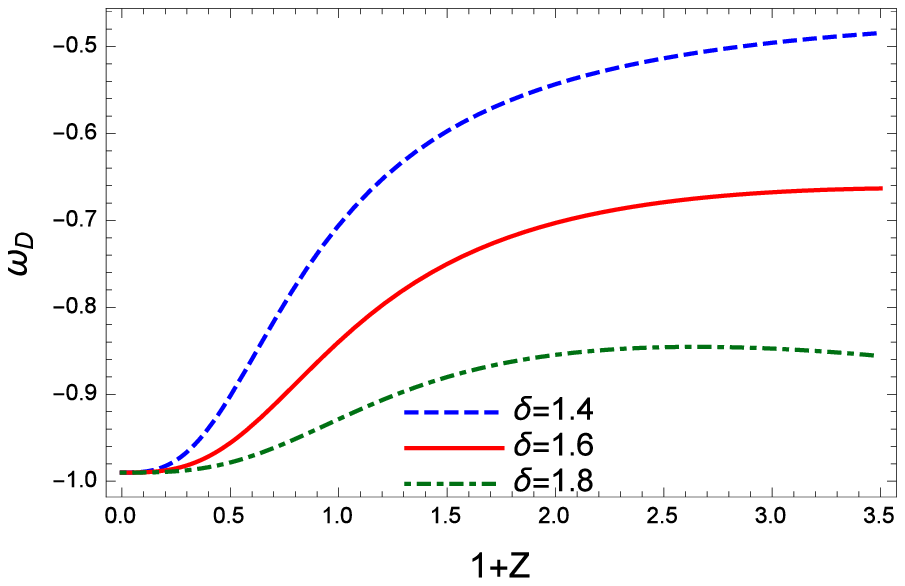}
\caption{$\omega_D(z)$ for
sign-changeable interacting THDE with Hubble radius as the IR cutoff.
}\label{w-z1}
\end{center}
\end{figure}

\begin{figure}[htp]
\begin{center}
\includegraphics[width=8cm]{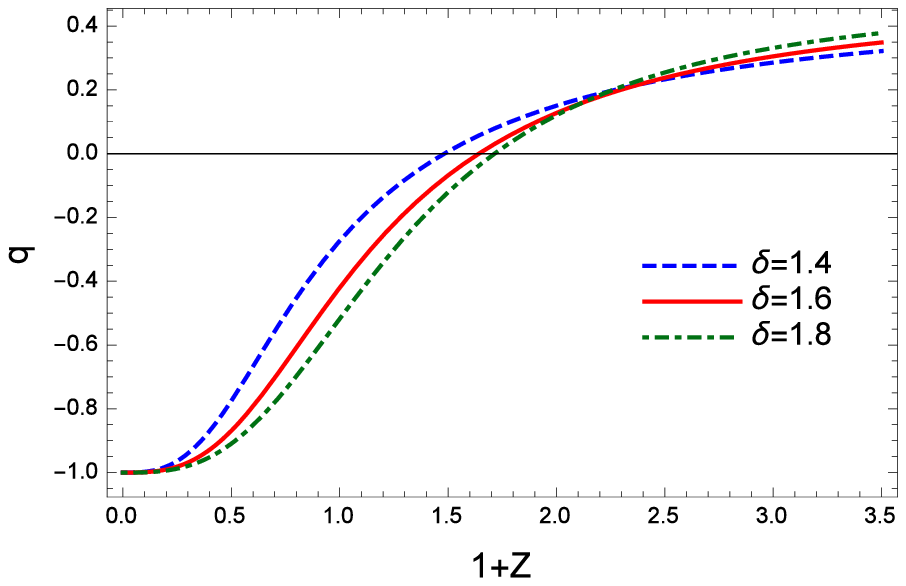}
\caption{$q(z)$ for sign-changeable interacting THDE with Hubble radius as the IR cutoff.
}\label{q-z1}
\end{center}
\end{figure}

\begin{figure}[htp]
\begin{center}
\includegraphics[width=8cm]{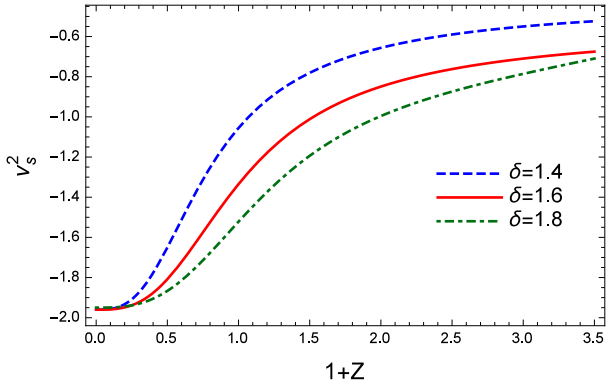}
\caption{$v_s^2 $ versus $z$ for sign-changeable interacting THDE with Hubble radius as the IR cutoff.
}\label{v-z1}
\end{center}
\end{figure}

\begin{figure}[htp]
\begin{center}
\includegraphics[width=8cm]{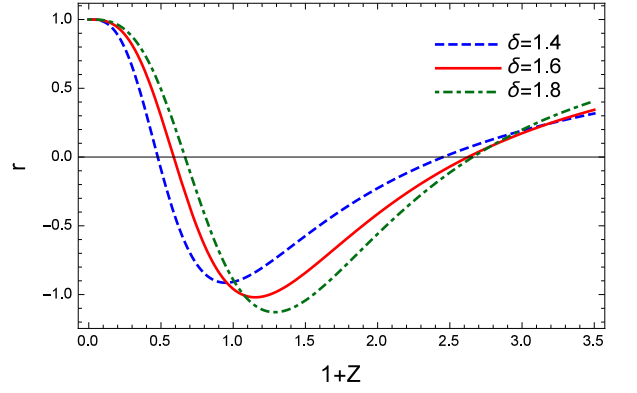}
\caption{$r(z)$ for sign-changeable interacting THDE with Hubble radius as the IR cutoff.
}\label{r-z1}
\end{center}
\end{figure}

\section{Sign-changeable interacting THDE with HUBBLE RADIUS AS IR CUTOFF IN BI MODEL}
\begin{figure}[htp]
\begin{center}
\includegraphics[width=8cm]{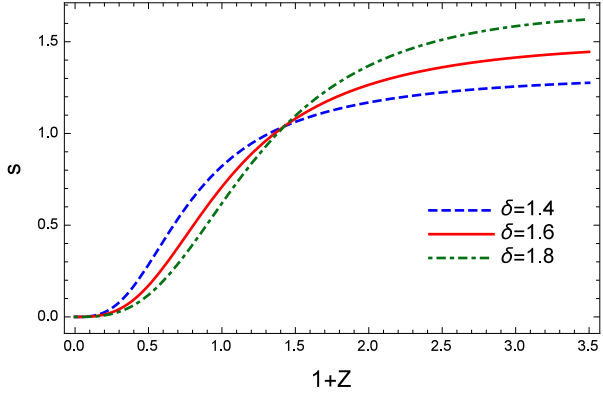}
\caption{The evolution of the statefinder parameter $s$ against $z$ for sign-changeable interacting THDE with Hubble radius as the IR cutoff.
}\label{s-z1}
\end{center}
\end{figure}

\begin{figure}[htp]
\begin{center}
\includegraphics[width=6cm]{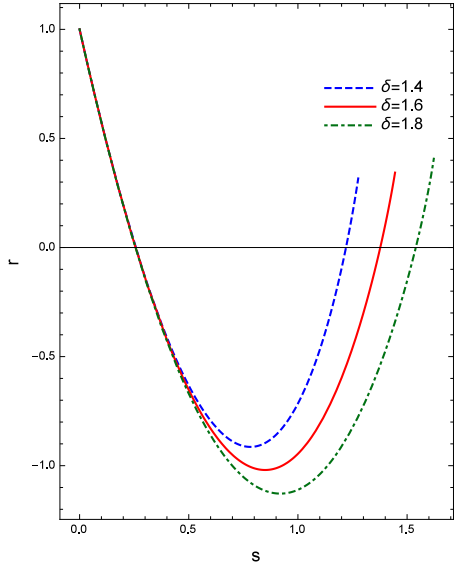}
\caption{The evolution of $r$ versus $s$ for sign-changeable interacting THDE with Hubble radius as the IR cutoff.
}\label{rs-z1}
\end{center}
\end{figure}

\begin{figure}[htp]
\begin{center}
\includegraphics[width=6cm]{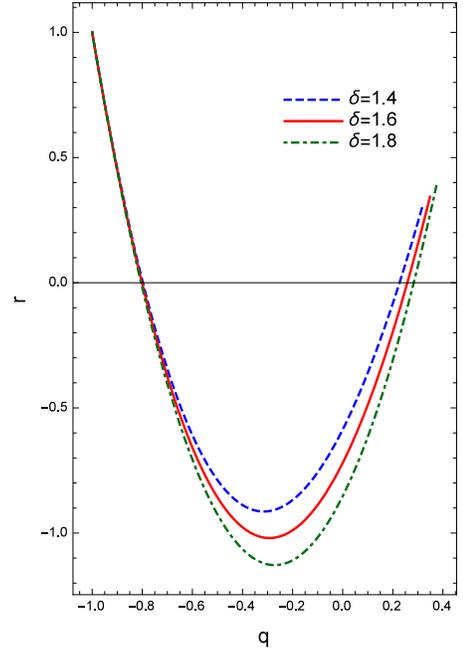}
\caption{$r$ against $q$ for sign-changeable interacting THDE with Hubble radius as the IR cutoff.
}\label{rq-z1}
\end{center}
\end{figure}

\begin{figure}[htp]
\begin{center}
\includegraphics[width=8cm]{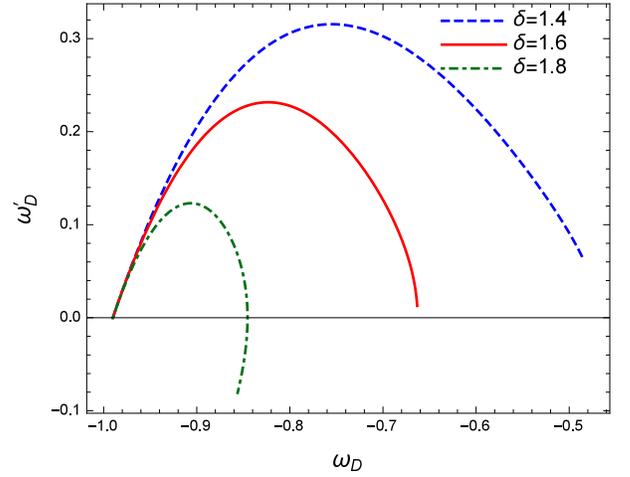}
\caption{The $\omega_D-{\omega}^{\prime}_{D}$ diagram for sign-changeable interacting THDE with Hubble radius as the IR cutoff. Here, we have taken
$\Omega_D(z=0)=0.73$, $H(z=0)=67$, $b^2=.01$ and $\Omega_{\sigma}=.001$}\label{ww-z1}
\end{center}
\end{figure}

THDE is defined as \cite{THDE}
\begin{eqnarray}\label{Trho}
\rho_D=BL^{2\delta-4},
\end{eqnarray}
leading to
\begin{eqnarray}\label{Hrho}
\rho_D=BH^{-2\delta+4},
\end{eqnarray}
where $B$ is an unknown parameter, if we consider the Hubble
radius $H^{-1}$ as the IR cutoff $L$. The time derivative of
relation (\ref{Hrho}) leads to
\begin{eqnarray}\label{dotHrho}
\dot{\rho}_{D}=\rho_D(-2\delta+4)\frac{\dot{H}}{H},
\end{eqnarray}
combined with the time derivative of the first Friedmann equation
and Eq.~(\ref{conm}) to reach
\begin{eqnarray}\label{dotHH}
\frac{\dot{H}}{H^{2}}= \frac{-3\Omega_D(b^2(1+u)+u)}{2-2\Omega_{\sigma}+\Omega_D(3b^2(1+u)-(-2\delta+4))},
\end{eqnarray}
where
\begin{eqnarray}\label{ratio}
u=\frac{\rho_m}{\rho_D}=\frac{\Omega_m}{\Omega_D}=-1+\frac{1}{\Omega_D}(1-\Omega_{\sigma}).
\end{eqnarray}
By inserting equation (\ref{dotHH}) in relation
(\ref{deceleration}), the deceleration parameter $q$ is found out
\begin{eqnarray}\label{qH}
q=-1+ \frac{3\Omega_D(b^2(1+u)+u)}{2-2\Omega_{\sigma}+\Omega_D(3b^2(1+u)-(-2\delta+4))}.
\end{eqnarray}
We can also find the EoS parameter, by combining Eqs.~(\ref{dotHrho}),~(\ref{conD}) and~(\ref{qH}) as
\begin{eqnarray}\label{EoSH}
\omega_D=-1-b^2 q(1+u)+(1+q)(\frac{-2\delta+4}{3}).
\end{eqnarray}
With the help of Eqs.~(\ref{Omega}),~(\ref{dotHrho}) and the
$\dot\Omega_{D}={\Omega}^{\prime}_{D} H$ relation, one reaches
\begin{eqnarray}\label{OmegaH}
\Omega_{D}^{\prime}=-2\Omega_D(-\delta+1)(1+q),
\end{eqnarray}
in which prime denotes derivative with respect to $\ln(a)$. Using the time derivative of Eqs.(\ref{qH}) and (\ref{EoSH}), we can
find $v_{s}^{2}$ and the ($r,s$) pair as
the statefinder parameters. Since these expressions are too long, we
do not present them here, study the evolution of these quantities
via figures. The model behavior has been depicted in Figs.~(\ref{Omega-z1}-\ref{ww-z1}) for the initial conditions $\Omega_D(z=0)=0.73$, $H(z=0)=67$, $b^2=.01$ and $\Omega_{\sigma}=.001$. These figures indicate that although the model is unstable at the classical level ($v_{s}^{2}<0$), it can cover the current accelerated universe for $\omega_D\geq-1$.

\section{Sign-changeable interacting THDE with Event horizon AS IR CUTOFF IN BI MODEL}

\begin{figure}[htp]
\begin{center}
\includegraphics[width=8cm]{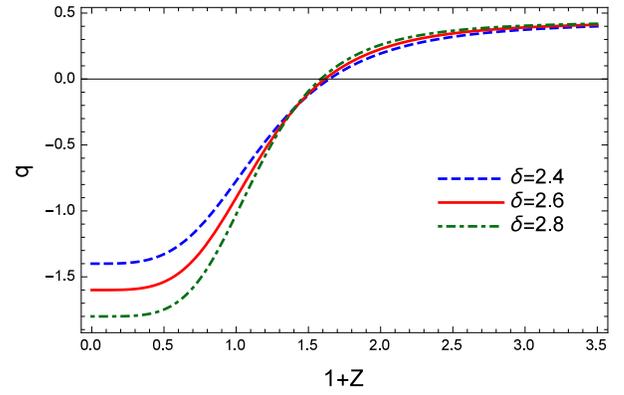}
\caption{The evolution of the deceleration parameter $q$ versus $z$ for sign-changeable interacting THDE with event horizon as IR cutoff.}\label{q-z2}
\end{center}
\end{figure}

\begin{figure}[htp]
\begin{center}
\includegraphics[width=8cm]{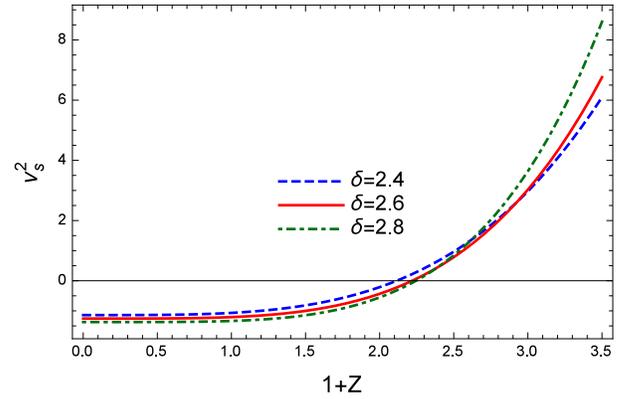}
\caption{$v_s^2(z)$ for sign-changeable interacting THDE with event horizon as IR cutoff.}\label{v-z2}
\end{center}
\end{figure}

\begin{figure}[htp]
\begin{center}
\includegraphics[width=8cm]{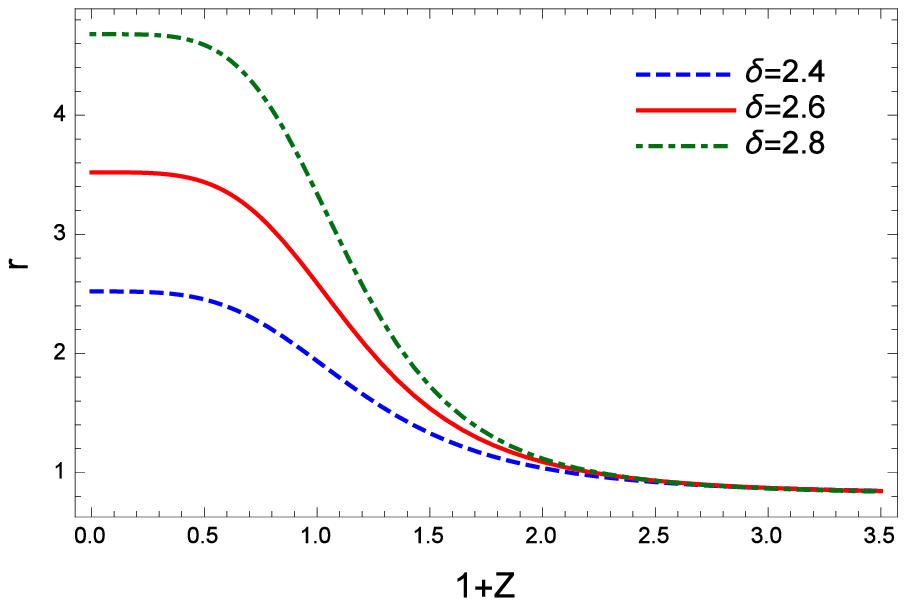}
\caption{The evolution of $r$ against $z$ for sign-changeable interacting THDE with event horizon as IR cutoff.}\label{r-z2}
\end{center}
\end{figure}

\begin{figure}[htp]
\begin{center}
\includegraphics[width=8cm]{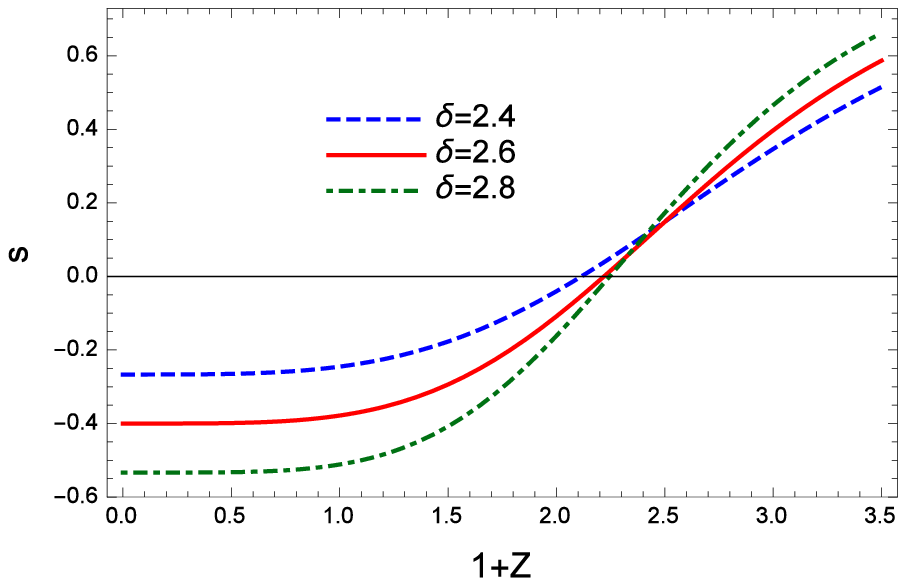}
\caption{The evolution of $s$ against $z$ for sign-changeable interacting THDE with event horizon as IR cutoff.}\label{s-z2}
\end{center}
\end{figure}

\begin{figure}[htp]
\begin{center}
\includegraphics[width=6cm]{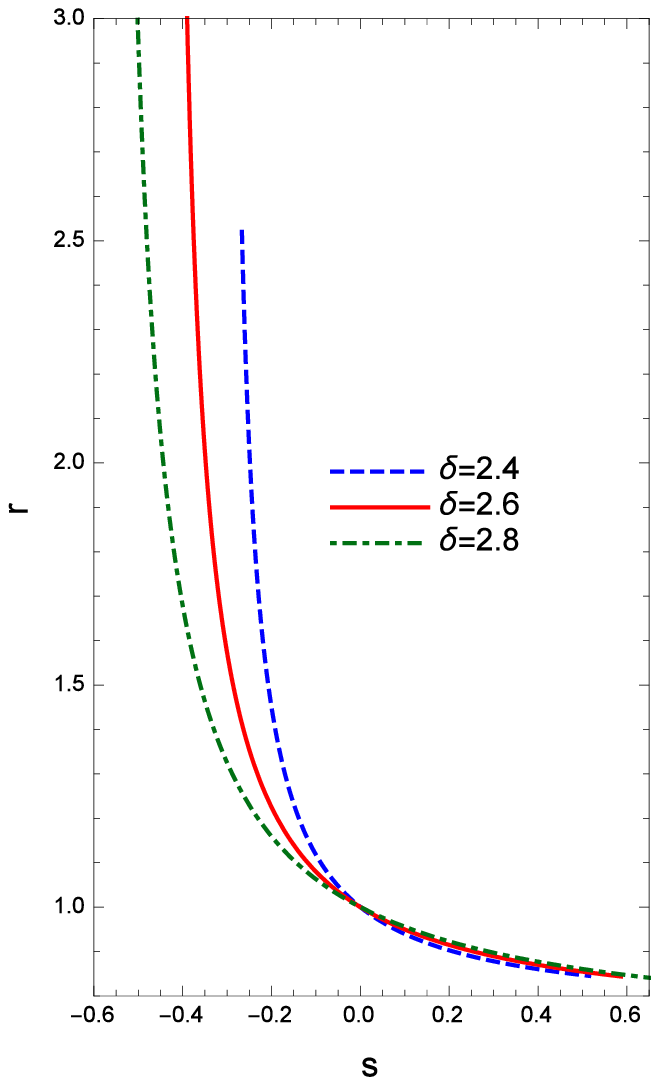}
\caption{The evolution of $r$ versus $s$ for sign-changeable interacting THDE with event horizon as IR cutoff.}\label{rs-z2}
\end{center}
\end{figure}

\begin{figure}[htp]
\begin{center}
\includegraphics[width=6cm]{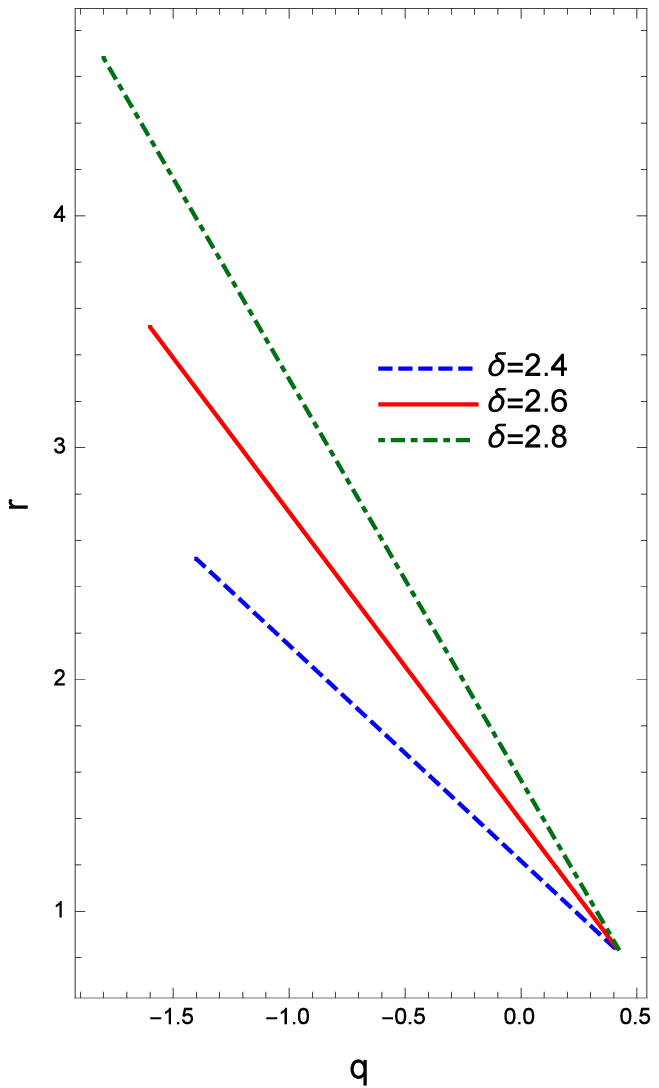}
\caption{The evolution of $r$ versus $q$ for sign-changeable interacting THDE with event horizon as IR cutoff.}\label{rq-z2}
\end{center}
\end{figure}

\begin{figure}[htp]
\begin{center}
\includegraphics[width=5cm]{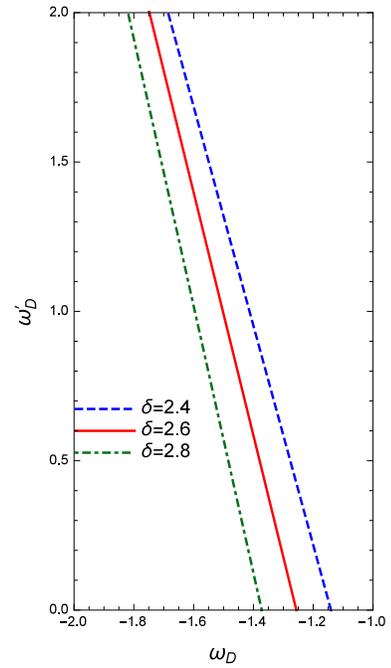}
\caption{The $\omega_D-{\omega}^{\prime}_{D}$ diagram for sign-changeable interacting THDE with event horizon as IR cutoff.}\label{ww-z2}
\end{center}
\end{figure}

If the future event horizon, defined as
\begin{equation}
R_{h}=a(t)\int_{t}^{\infty}{\frac{dt}{a(t)}},
\end{equation}
leading to
\begin{equation}\label{event}
\dot{R}_{h}=H R_h -1,
\end{equation}
is used as the IR cutoff ($L=R_h$), then
\begin{eqnarray}\label{Frho}
\rho_D=B R_{h}^{2\delta-4}.
\end{eqnarray}
By combining the time derivative of the above equation with
Eq.(\ref{event}), we find
\begin{eqnarray}\label{dotFrho}
\dot{\rho}_{D}=\rho_D(2\delta-4)H(1-F),
\end{eqnarray}
where $F=(\frac{3\Omega_D
H^{2\delta-2}}{B})^{\frac{1}{-2\delta+4}}$.
\begin{figure}[htp]
\begin{center}
\includegraphics[width=8cm]{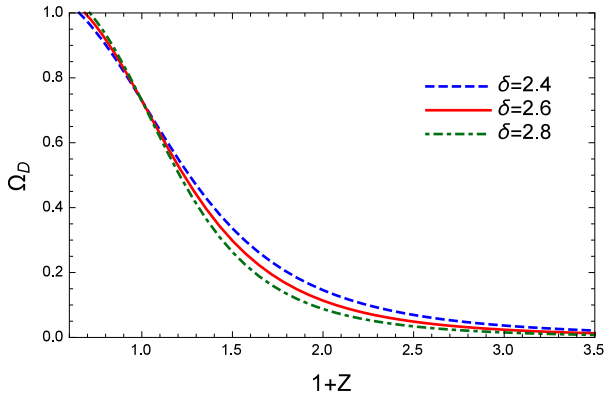}
\caption{The evolution of $\Omega_D$ versus redshift parameter $z$ for
sign-changeable interacting THDE with event horizon as IR cutoff in BI model.}\label{Omega-z2}
\end{center}
\end{figure}
Inserting Eq.(\ref{dotFrho}) in Eq.(\ref{conD}), we easily reach at
\begin{eqnarray}\label{EoSev}
\omega_D=-1 -b^2 q(1+u)-(\frac{2\delta-4}{3})(1-F).
\end{eqnarray}
\begin{figure}[htp]
\begin{center}
\includegraphics[width=8cm]{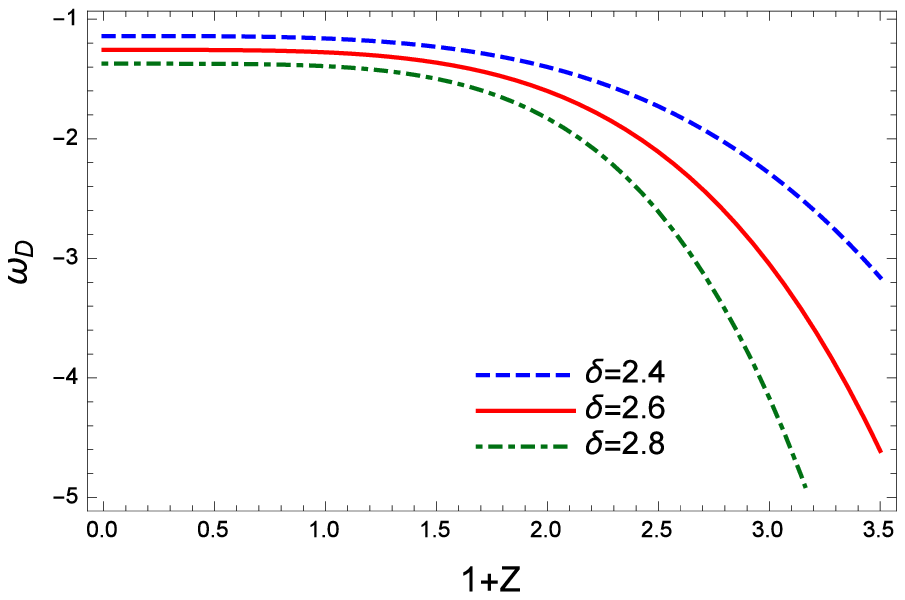}
\caption{The evolution of $\omega_D$ versus $z$ for
sign-changeable interacting THDE with event horizon as IR cutoff.}\label{w-z2}
\end{center}
\end{figure}
Additionally, combining the time derivative of Eq.(\ref{ani1}) with Eqs.(\ref{dotFrho})
and (\ref{conm}), one obtains
\begin{eqnarray}\label{dotHev}
\frac{\dot{H}}{H^{2}}=\frac{\Omega_D(-3b^2 (1+u)-3u+(2\delta-4)(1-F)}{2-2\Omega_{\sigma}+3b^2 \Omega_D(1+u)},
\end{eqnarray}
and
\begin{eqnarray}\label{qev}
q=-1-\frac{\Omega_D(-3b^2 (1+u)-3u+(2\delta-4)(1-F))}{2-2\Omega_{\sigma}+3b^2 \Omega_D(1+u)}.
\end{eqnarray}
Using the $\Omega_D$ expression and Eq.(\ref{dotFrho}), we find
\begin{eqnarray}\label{Omegaev}
\Omega_{D}^{\prime}=\Omega_D((2\delta-4)(1-F)+2(1+q)).
\end{eqnarray}
Although $v_{s}^{2}$, $r$ and $s$ can be found out by by taking the time derivative of Eqs.(\ref{EoSev}) and
(\ref{qev}), since they are too long relations, we do not present them here, and only plot them. In Figs.~(\ref{Omega-z2}-\ref{ww-z2}), the model behavior is depicted for the initial conditions $\Omega_D(z=0)=0.73$, $H(z=0)=67$, $b^2=.1$, $B=2.4$ and $\Omega_{\sigma}=.001$ indicating that, unlike the previous case, the phantom behaviors is unavoidable.
\section{Sign-changeable interacting THDE with Particle horizon AS IR CUTOFF IN BI MODEL}
\begin{figure}[htp]
\begin{center}
\includegraphics[width=8cm]{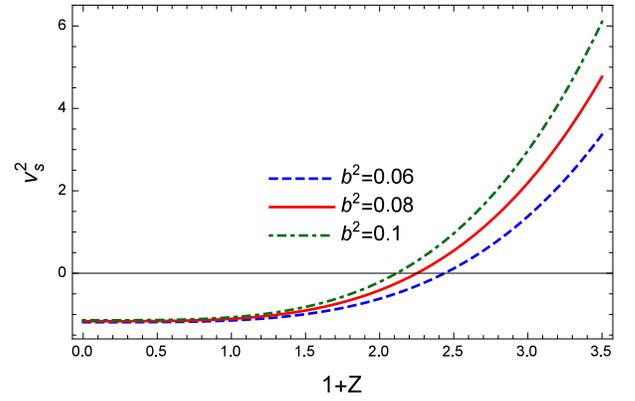}
\caption{The evolution of  the squared of sound speed $v_s^2 $ versus redshift parameter $z$ for sign-changeable interacting THDE with particle horizon as IR cutoff.}\label{v-z3}
\end{center}
\end{figure}

\begin{figure}[htp]
\begin{center}
\includegraphics[width=8cm]{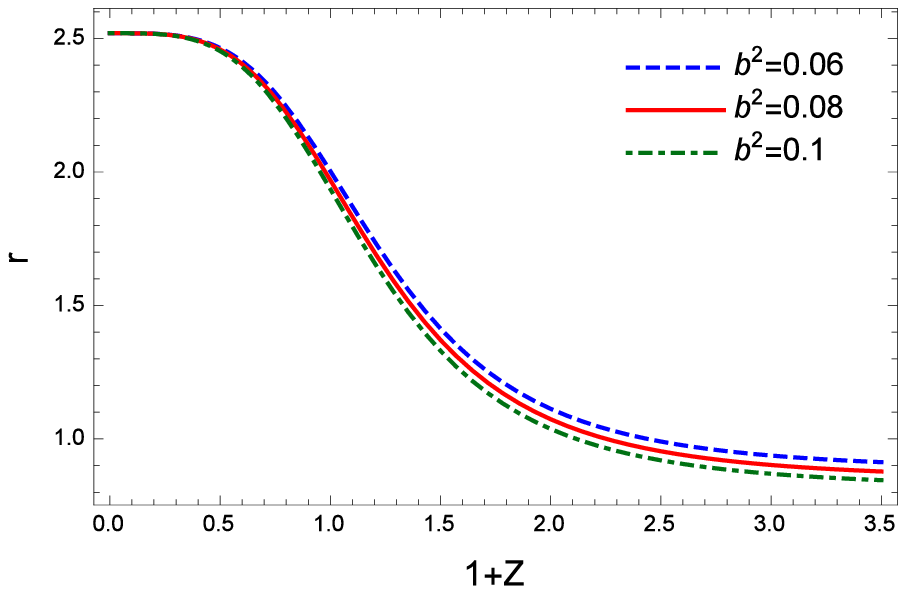}
\caption{The evolution of the statefinder parameter $r$ versus the redshift parameter $z$ for sign-changeable interacting THDE with particle horizon as IR cutoff.}\label{r-z3}
\end{center}
\end{figure}

\begin{figure}[htp]
\begin{center}
\includegraphics[width=8cm]{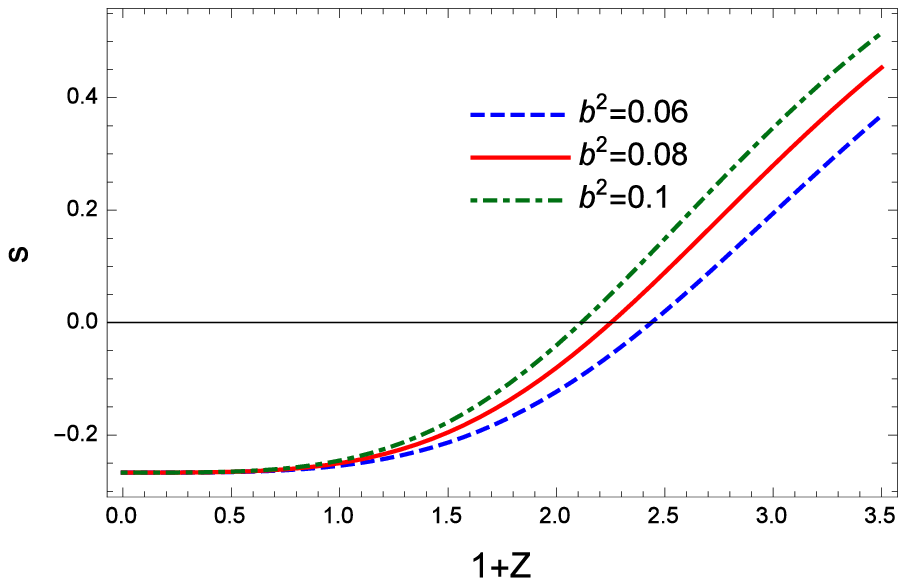}
\caption{The evolution of the statefinder parameter $s$ versus the redshift parameter $z$ for sign-changeable interacting THDE with particle horizon as IR cutoff.}\label{s-z3}
\end{center}
\end{figure}

\begin{figure}[htp]
\begin{center}
\includegraphics[width=7cm]{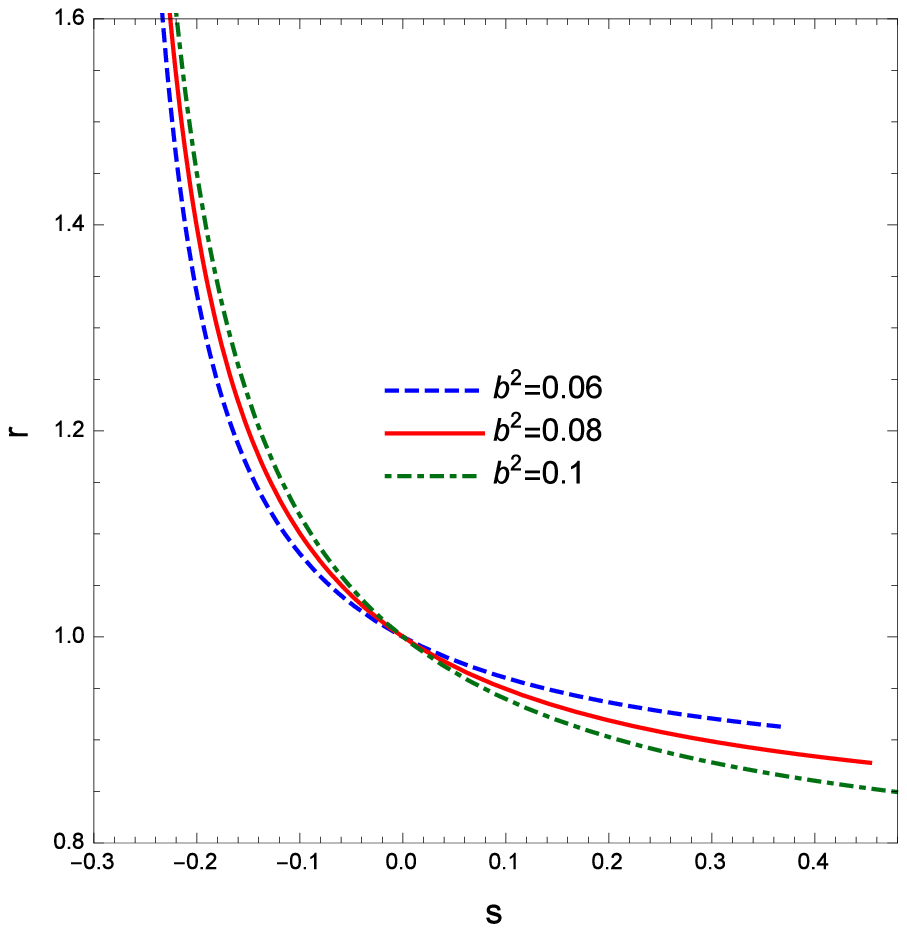}
\caption{The evolution of the statefinder parameter $r$ versus $s$ for sign-changeable interacting THDE with particle horizon as IR cutoff.}\label{rs-z3}
\end{center}
\end{figure}

\begin{figure}[htp]
\begin{center}
\includegraphics[width=7cm]{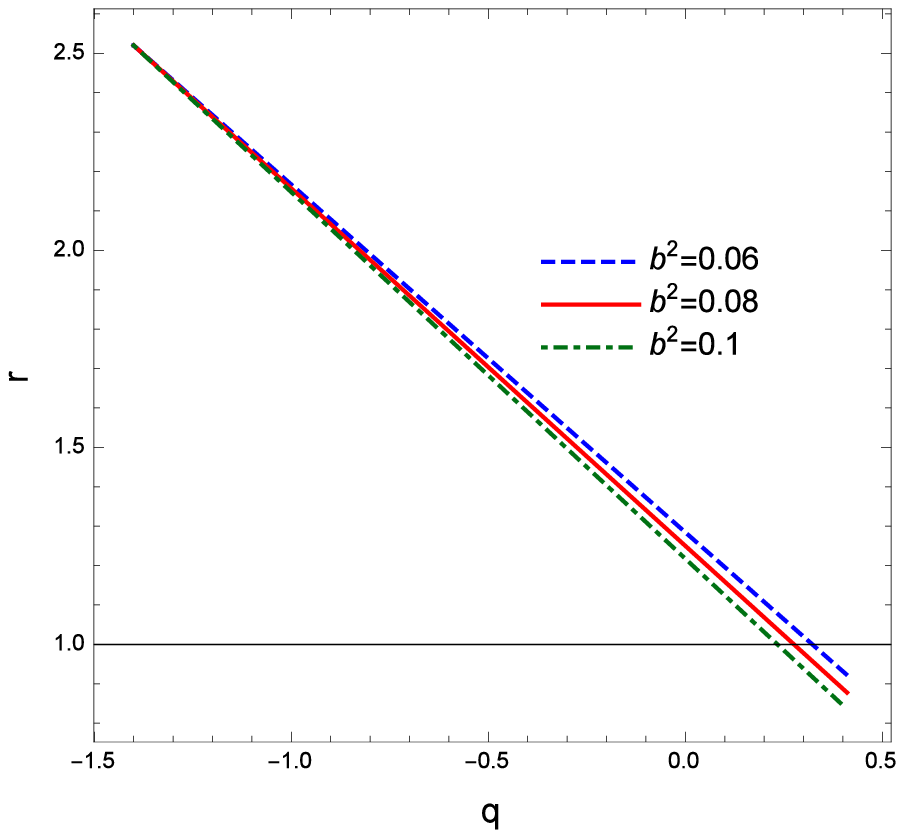}
\caption{The evolution of the statefinder parameter $r$ versus the deceleration parameter $q$ for sign-changeable interacting THDE with particle horizon as IR cutoff.}\label{rq-z3}
\end{center}
\end{figure}

\begin{figure}[htp]
\begin{center}
\includegraphics[width=5cm]{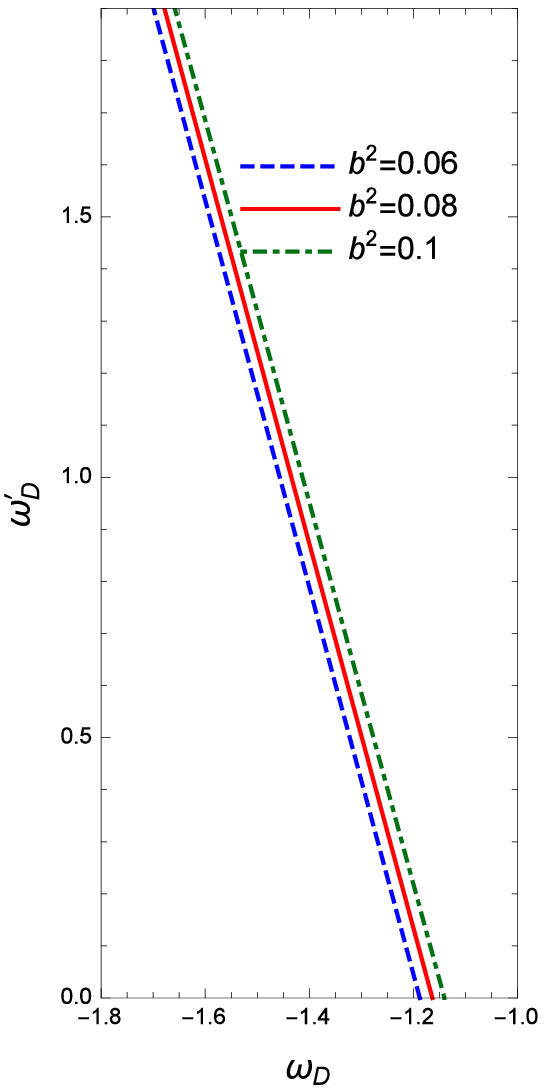}
\caption{The $\omega_D-{\omega}^{\prime}_{D}$ diagram for sign-changeable interacting THDE with particle horizon as IR cutoff.}\label{ww-z3}
\end{center}
\end{figure}

The particle horizon, introduced
as \cite{Li2004}
\begin{equation}\label{particle}
\dot{R}_{p}=H R_p +1,
\end{equation}
is considered as the IR cutoff in this section. In this manner, Eq.(\ref{Trho}) leads to
\begin{eqnarray}\label{Prho}
\rho_D=B R_{p}^{2\delta-4},
\end{eqnarray}
and
\begin{eqnarray}\label{dotPrho}
\dot{\rho}_{D}=\rho_D(2\delta-4)H(1+F),
\end{eqnarray}
the time derivative of $\rho_D$.
\begin{figure}[htp]
\begin{center}
\includegraphics[width=8cm]{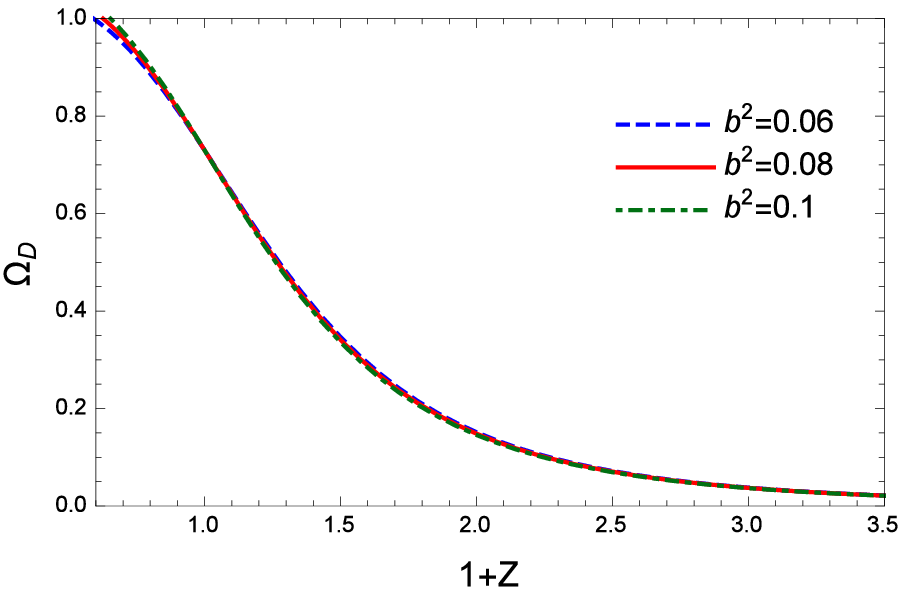}
\caption{The evolution of $\Omega_D$ versus redshift parameter $z$ for
sign-changeable interacting THDE with particle horizon as IR cutoff in.}\label{Omega-z3}
\end{center}
\end{figure}
Combining Eq.(\ref{dotPrho}) with the conservation law (\ref{conD}), one obtains
\begin{eqnarray}\label{EoSP}
\omega_D=-1 -b^2 q(1+u)-(\frac{2\delta-4}{3})(1+F).
\end{eqnarray}
In addition, by using Eqs.~(\ref{ani1}),~(\ref{dotPrho}) and~(\ref{deceleration}), the deceleration parameter $q$ is found out as
\begin{eqnarray}\label{qp}
q=-1-\frac{\Omega_D(-3b^2 (1+u)-3u+(2\delta-4)(1+F))}{2-2\Omega_{\sigma}+3b^2 \Omega_D(1+u)}.
\end{eqnarray}
\begin{figure}[htp]
\begin{center}
\includegraphics[width=8cm]{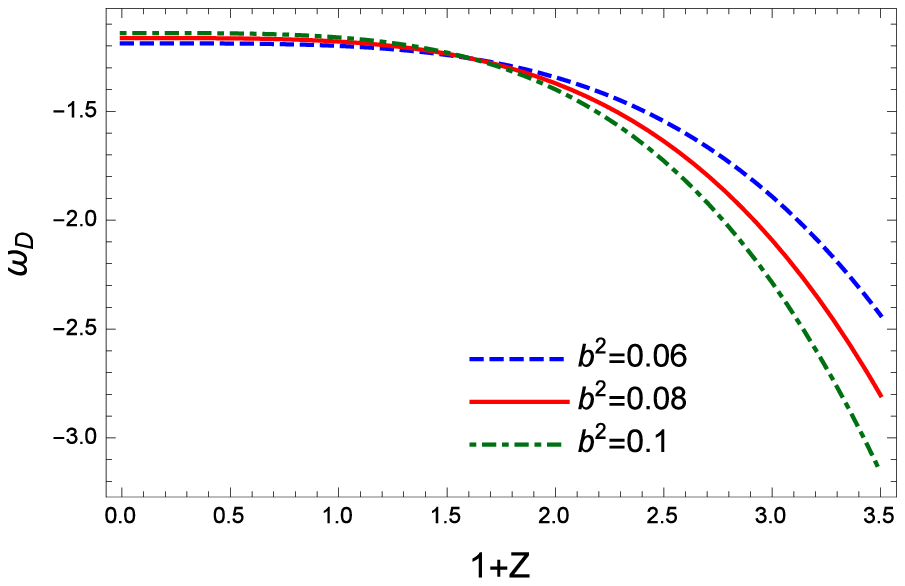}
\caption{The evolution of $\omega_D$ versus redshift parameter $z$ for
sign-changeable interacting THDE with particle horizon as IR cutoff in.}\label{w-z3}
\end{center}
\end{figure}
Moreover, one can insert $\Omega_{D}=\frac{\rho_{D}}{3m_{p}^{2}H^{2}}$ in
Eq.(\ref{dotPrho}) to reach at
\begin{eqnarray}\label{Omegap}
\Omega_{D}^{\prime}=\Omega_D((2\delta-4)(1+F)+2(1+q)).
\end{eqnarray}
\begin{figure}[htp]
\begin{center}
\includegraphics[width=8cm]{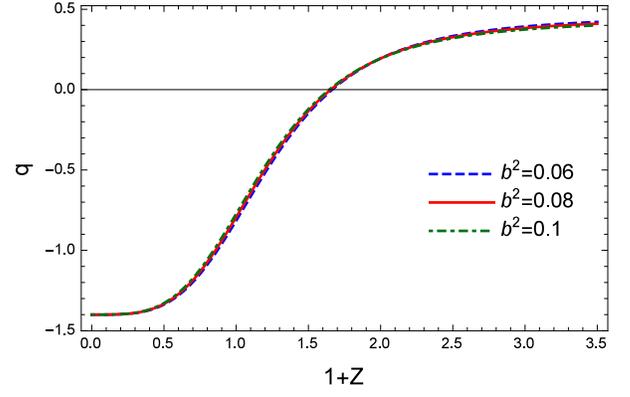}
\caption{The evolution of the deceleration parameter $q$ versus redshift parameter $z$ for sign-changeable interacting THDE with particle horizon as IR cutoff in.}\label{q-z3}
\end{center}
\end{figure}
The same as the previous sections, since the expressions of $v_{s}^{2}$, $r$ and $s$ are too long, we only plot them and do not write them here. In Figs. (\ref{Omega-z3}-\ref{ww-z3}), the system parameters have been plotted by employing the initial conditions $\Omega_D(z=0)=0.73$, $H(z=0)=67$, $\delta=2.4$, $B=2.4$ and $\Omega_{\sigma}=.001$.
\section{Sign-changeable interacting THDE with GO horizon AS IR CUTOFF IN BI MODEL}
\begin{figure}[htp]
\begin{center}
\includegraphics[width=8cm]{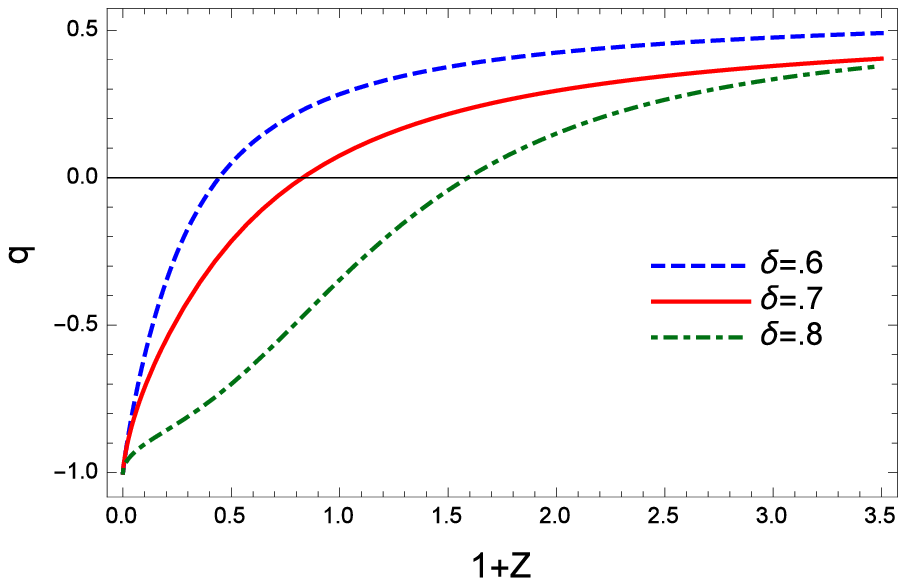}
\caption{The evolution of the deceleration parameter $q$ versus redshift parameter $z$ for sign-changeable interacting THDE with GO horizon as IR cutoff.}\label{q-z4}
\end{center}
\end{figure}

\begin{figure}[htp]
\begin{center}
\includegraphics[width=8cm]{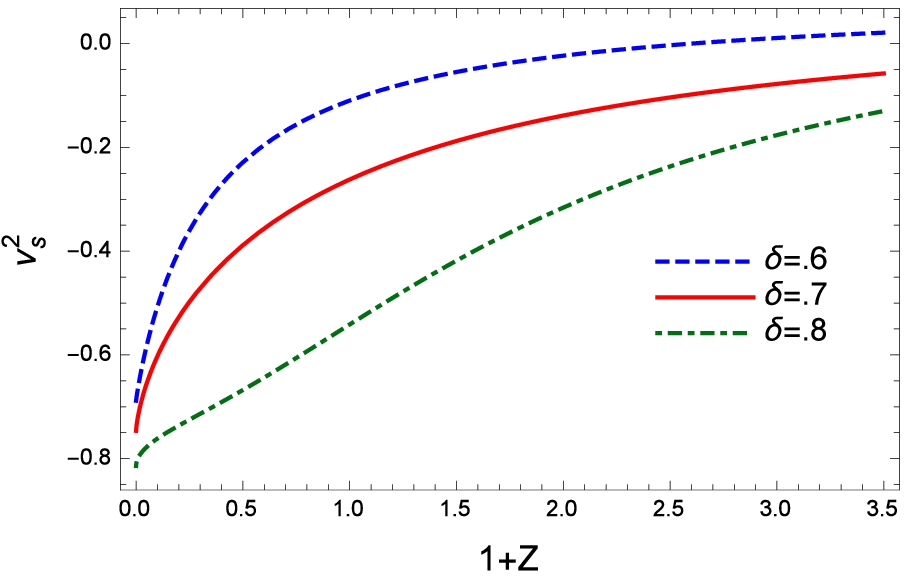}
\caption{The evolution of the squared of sound speed $v_s^2 $ versus redshift parameter $z$ for sign-changeable interacting THDE with GO horizon as IR cutoff.}\label{v-z4}
\end{center}
\end{figure}

\begin{figure}[htp]
\begin{center}
\includegraphics[width=7cm]{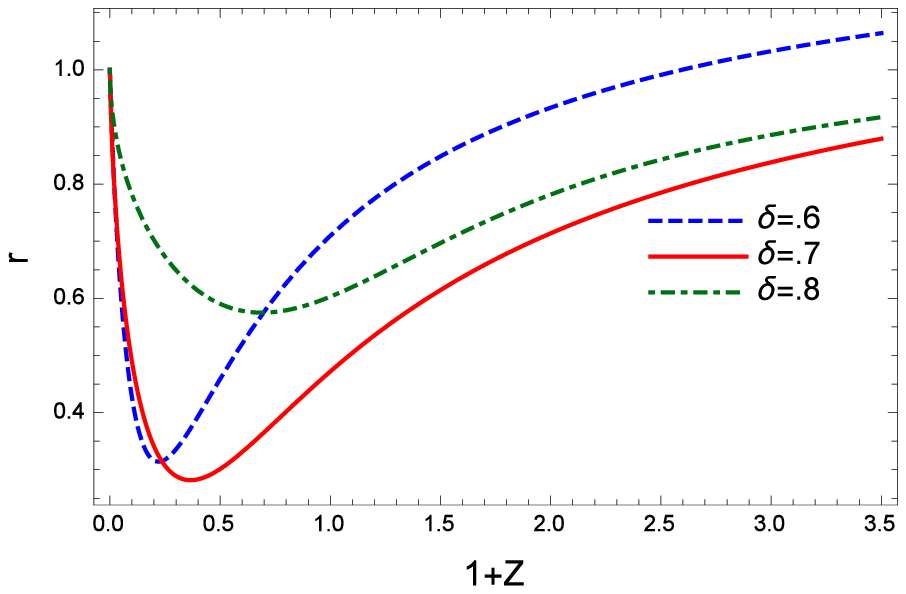}
\caption{The evolution of the statefinder parameter $r$ versus the redshift parameter $z$ for sign-changeable interacting THDE with GO horizon as IR cutoff.}\label{r-z4}
\end{center}
\end{figure}

\begin{figure}[htp]
\begin{center}
\includegraphics[width=7cm]{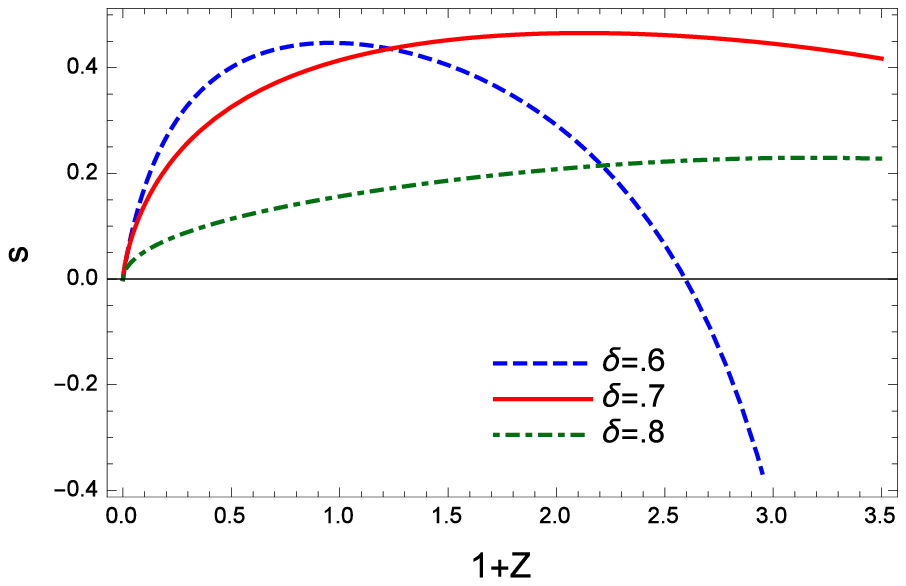}
\caption{The evolution of the statefinder parameter $s$ versus the redshift parameter $z$ for sign-changeable interacting THDE with GO horizon as IR cutoff.}\label{s-z4}
\end{center}
\end{figure}

\begin{figure}[htp]
\begin{center}
\includegraphics[width=7cm]{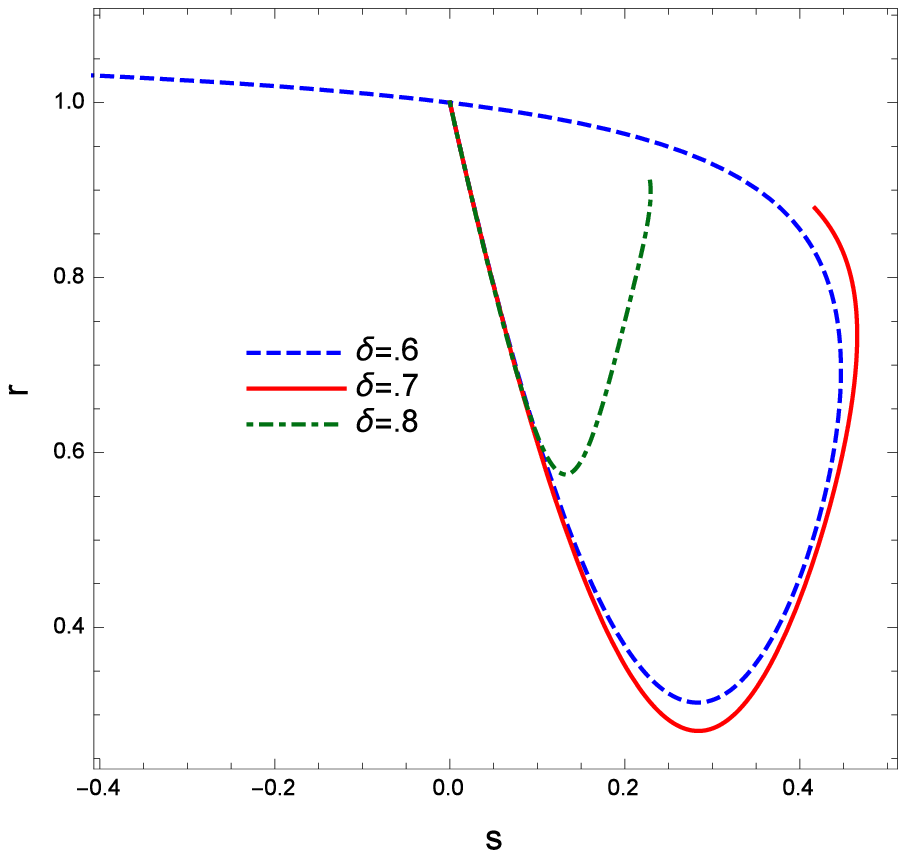}
\caption{The evolution of the statefinder parameter $r$ versus $s$ for sign-changeable interacting THDE with GO horizon as IR cutoff.}\label{rs-z4}
\end{center}
\end{figure}

\begin{figure}[htp]
\begin{center}
\includegraphics[width=8cm]{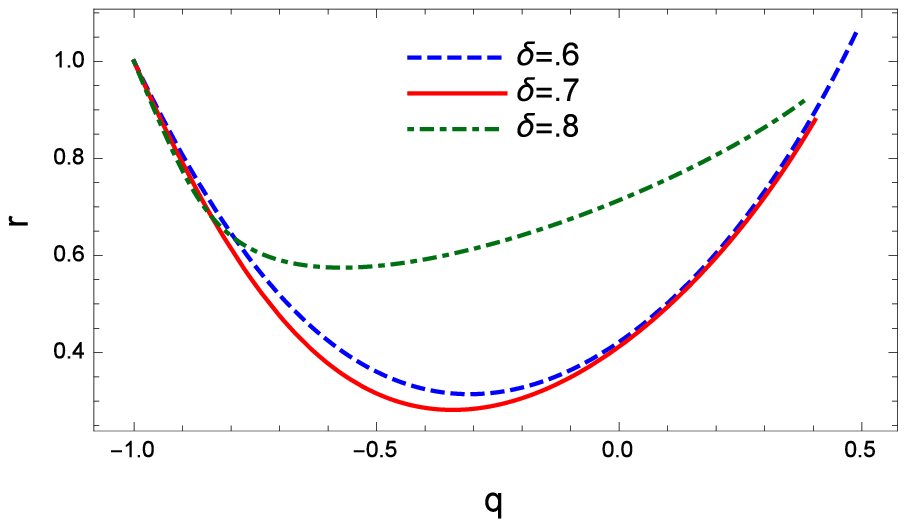}
\caption{The evolution of the statefinder parameter $r$ versus the deceleration parameter $q$ for sign-changeable interacting THDE with GO horizon as IR cutoff.}\label{rq-z4}
\end{center}
\end{figure}

\begin{figure}[htp]
\begin{center}
\includegraphics[width=8cm]{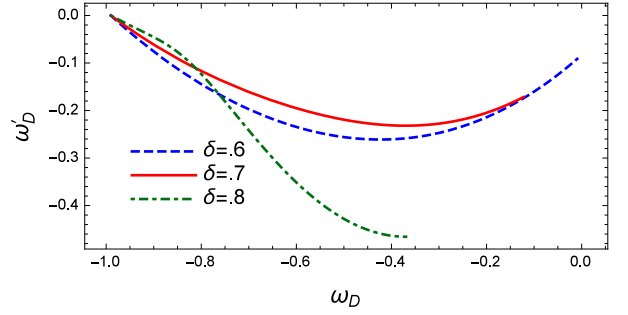}
\caption{The $\omega_D-{\omega}^{\prime}_{D}$ diagram for sign-changeable interacting THDE with GO horizon as IR cutoff.}\label{ww-z4}
\end{center}
\end{figure}

Bearing Eq.~(\ref{Trho}) in mind, and employing the GO cutoff \cite{Li20041,Li20042}, the energy density of THDE is given by
 \begin{equation}\label{GOrho}
\rho_D=(\alpha H^2+\beta \dot{H})^{-\delta +2},
\end{equation}
rewritten as
\begin{equation}\label{GOdotH}
\frac{\dot{H}}{H^2}= \frac{1}{\beta}\left(\frac{{(3m_p^2 \Omega_D)}^{\frac{1}{2-\delta}}}{{H}^{\frac{2-2\delta}{2-\delta}}}-\alpha \right).
\end{equation}
Now, by using relation (\ref{deceleration}), we have
\begin{equation}\label{qGO}
q=-1-\frac{1}{\beta}\left(\frac{{(3m_p^2
\Omega_D)}^{\frac{1}{2-\delta}}}{{H}^{\frac{2-2\delta}{2-\delta}}}-\alpha
\right).
\end{equation}
\begin{figure}[htp]
\begin{center}
\includegraphics[width=8cm]{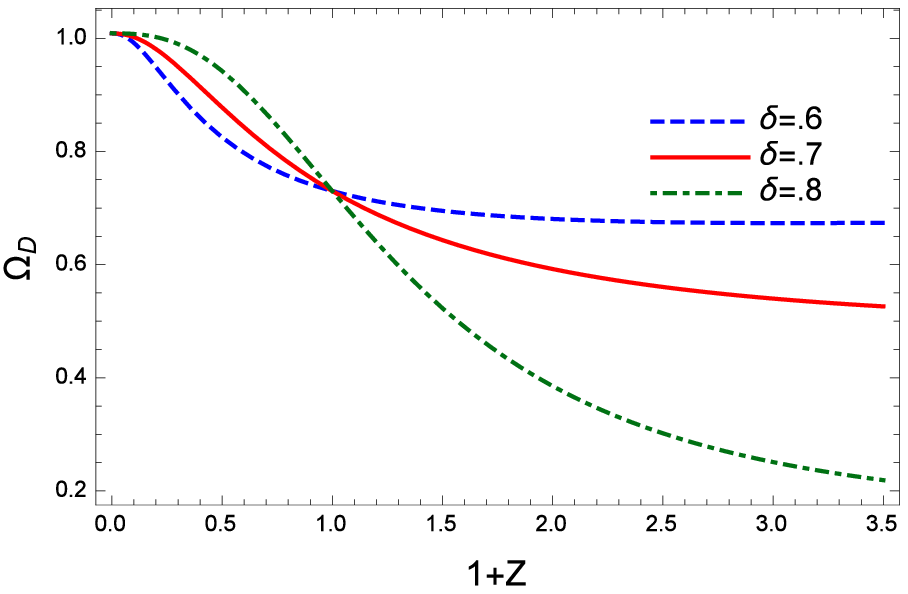}
\caption{The evolution of $\Omega_D$ versus redshift parameter $z$ for
sign-changeable interacting THDE with GO horizon as IR cutoff.}\label{Omega-z4}
\end{center}
\end{figure}
\begin{figure}[htp]
\begin{center}
\includegraphics[width=8cm]{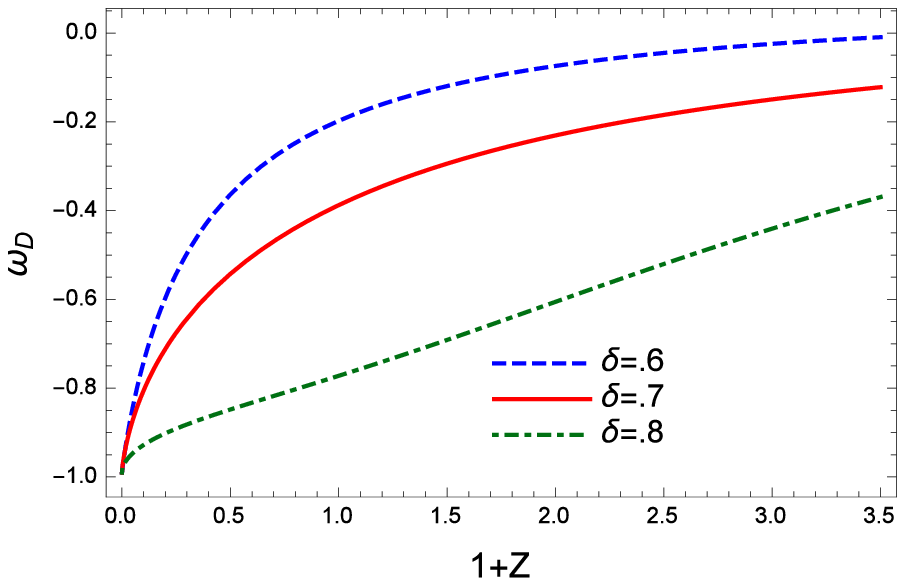}
\caption{The evolution of $\omega_D$ versus redshift parameter $z$ for
sign-changeable interacting THDE with GO horizon as IR cutoff.}\label{w-z4}
\end{center}
\end{figure}
In addition, inserting the time derivative of Eq.(\ref{ani1}) in
Eq.(\ref{conm}), one obtains
\begin{eqnarray}\label{dotrhoGo}
&&\frac{\dot{\rho}_D}{3m_p^2H^3}=\frac{\dot{H}}{H^2}(2-2\Omega_{\sigma}+3 b^2 \Omega_D(1+u))\\&&+3b^2 \Omega_D(1+u)+3(1-\Omega_{\sigma}-\Omega_D)\nonumber,
\end{eqnarray}
combined with
\begin{equation}\label{GoOmega1}
\dot{\Omega}_D=\frac{\dot{\rho}_D}{3M_p^2H^2}-2\Omega_D\frac{\dot{H}}{H},
\end{equation}
to reach at
\begin{eqnarray}\label{GoOmega2}
&&{\Omega}^\prime_D=(3b^2 \Omega_D(1+u) +3(1-\Omega_{\sigma}-\Omega_D)\\&&-(1+q)(2-2\Omega_{\sigma}-2\Omega_D+3b^2\Omega_D(1+u))).
\end{eqnarray}
The EoS parameter $\omega_D$) of THDE is found out as
\begin{equation}\label{EoSGo}
\omega_D=-1-\frac{1}{3\Omega_D}(3(1-\Omega_{\sigma}-\Omega_D)-(1+q)(2-2\Omega_D)),
\end{equation}
by combining Eq.(\ref{dotrhoGo}) with Eq.(\ref{conD}). Figs.~(\ref{Omega-z4}-\ref{ww-z4}) show the behavior of the model parameters during the cosmic evolution by presuming $\Omega_D(z=0)=0.73$, $H(z=0)=67$, $\alpha=.8$,$\beta=.5$, $b^2=.01$ as the initial conditions.
\section{Sign-changeable interacting THDE with Ricci horizon AS IR CUTOFF IN BI MODEL}
Let us use the Ricci cutoff \cite{Gao} for describing cosmological
parameters. It leads to
\begin{equation}\label{Rrho}
\rho_D=\lambda(2H^2+ \dot{H})^{-\delta +2},
\end{equation}
where $\lambda$ is the unknown HDE constant as usual
\cite{Li2004,Gao}, for THDE.
\begin{figure}[htp]
\begin{center}
\includegraphics[width=8cm]{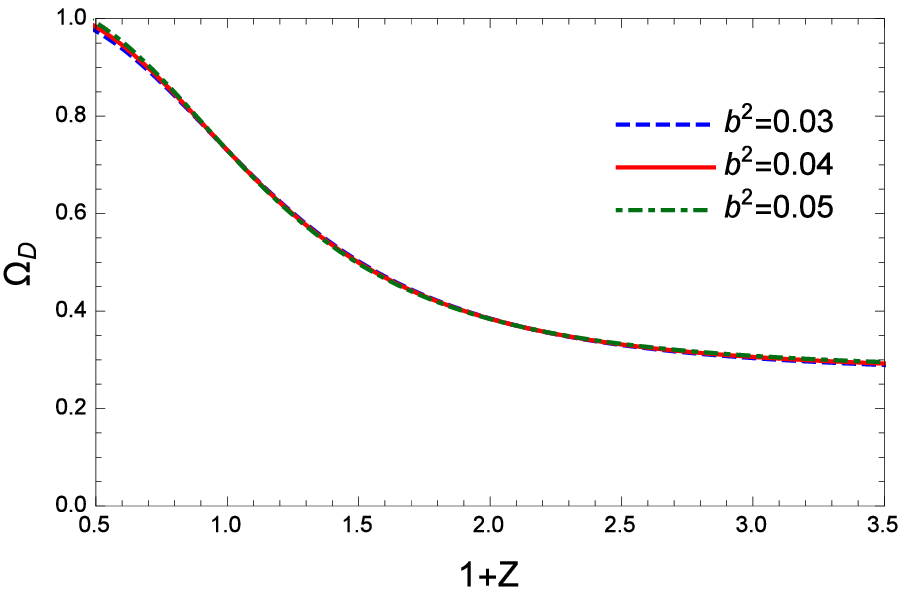}
\caption{The evolution of $\Omega_D$ versus redshift parameter $z$ for
sign-changeable interacting THDE with Ricci horizon.}\label{Omega-z5}
\end{center}
\end{figure}
By rewriting Eq.(\ref{Rrho}) as
\begin{equation}\label{RdotH}
\frac{\dot{H}}{H^2}= \left(\frac{{(3 {\lambda}^{-1} m_p^2 \Omega_D)}^{\frac{1}{2-\delta}}}{{H}^{\frac{2-2\delta}{2-\delta}}}-2 \right),
\end{equation}
and using relation (\ref{deceleration}), we get
\begin{equation}\label{qR}
q=-1-\left(\frac{{(3 {\lambda}^{-1} m_p^2 \Omega_D)}^{\frac{1}{2-\delta}}}{{H}^{\frac{2-2\delta}{2-\delta}}}-2 \right).
\end{equation}
\begin{figure}[htp]
\begin{center}
\includegraphics[width=8cm]{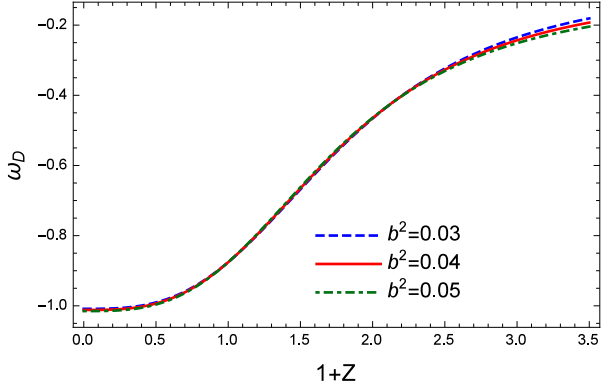}
\caption{The evolution of $\omega_D$ versus redshift parameter $z$ for
sign-changeable interacting THDE with Ricci horizon as IR cutoff in BI model. Here, we have taken
$\Omega_D(z=0)=0.73$, $H(z=0)=67$, $\delta=1$, $\lambda=1.5$  and $\Omega_{\sigma}=.001$}\label{w-z5}
\end{center}
\end{figure}
Moreover, combining Eqs.(\ref{dotrhoGo}) with (\ref{GoOmega1}), one finds
\begin{eqnarray}\label{OmegaR}
&&{\Omega}^\prime_D=(3b^2 \Omega_D(1+u) +3(1-\Omega_{\sigma}-\Omega_D)\nonumber\\&&-(1+q)(2-2\Omega_{\sigma}-2\Omega_D+3b^2\Omega_D(1+u))).
\end{eqnarray}
\begin{figure}[htp]
\begin{center}
\includegraphics[width=8cm]{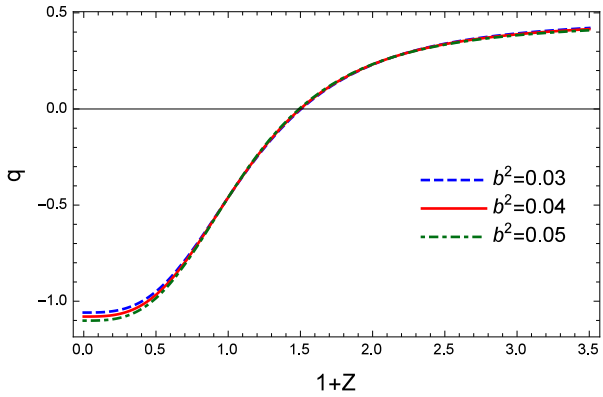}
\caption{The evolution of the deceleration parameter $q$ versus redshift parameter $z$ for sign-changeable interacting THDE with Ricci horizon as IR cutoff.}\label{q-z5}
\end{center}
\end{figure}
Following the recipe of the previous section, the EoS parameter is calculated as
\begin{equation}\label{EoSR}
\omega_D=-1-\frac{1}{3\Omega_D}(3(1-\Omega_{\sigma}-\Omega_D)-(1+q)(2-2\Omega_D)).
\end{equation}
\begin{figure}[htp]
\begin{center}
\includegraphics[width=8cm]{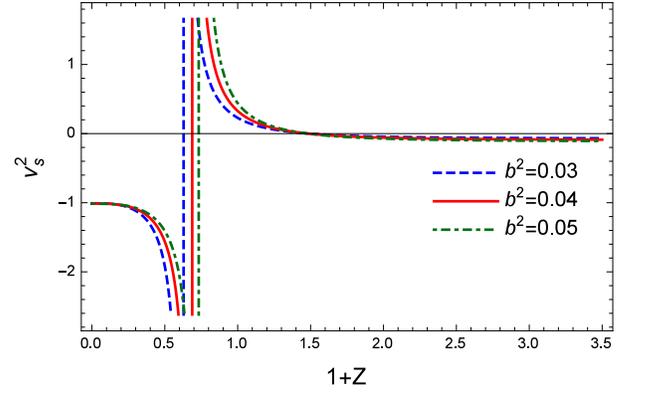}
\caption{The evolution of  the squared of sound speed $v_s^2 $ versus redshift parameter $z$ for sign-changeable interacting THDE with Ricci horizon as IR cutoff.}\label{v-z5}
\end{center}
\end{figure}
In Fig.~(\ref{Omega-z5}-\ref{w-z5}) the model parameters have been plotted for the initial conditions $\Omega_D(z=0)=0.73$, $H(z=0)=67$, $\delta=1$, $\lambda=1.5$ and $\Omega_{\sigma}=.001$ and $\Omega_{\sigma}=.001$.
\begin{figure}[htp]
\begin{center}
\includegraphics[width=8cm]{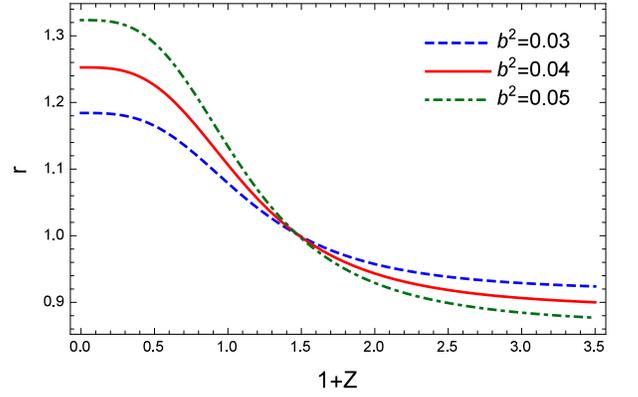}
\caption{The evolution of the statefinder parameter $r$ versus the redshift parameter $z$ for sign-changeable interacting THDE with Ricci horizon as IR cutoff.}\label{r-z5}
\end{center}
\end{figure}
\begin{figure}[htp]
\begin{center}
\includegraphics[width=8cm]{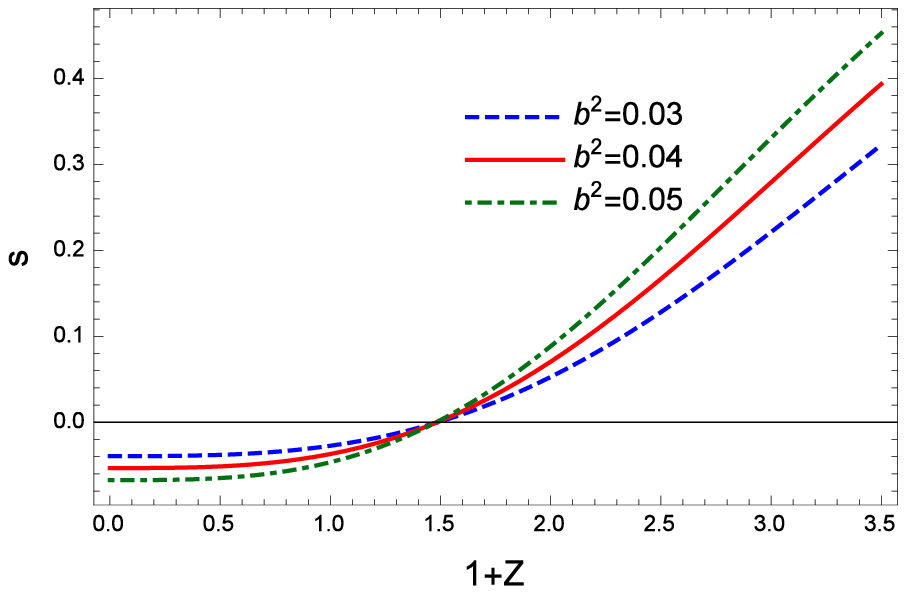}
\caption{The evolution of the statefinder parameter $s$ versus the redshift parameter $z$ for sign-changeable interacting THDE with Ricci horizon as IR cutoff.}\label{s-z5}
\end{center}
\end{figure}
\begin{figure}[htp]
\begin{center}
\includegraphics[width=6cm]{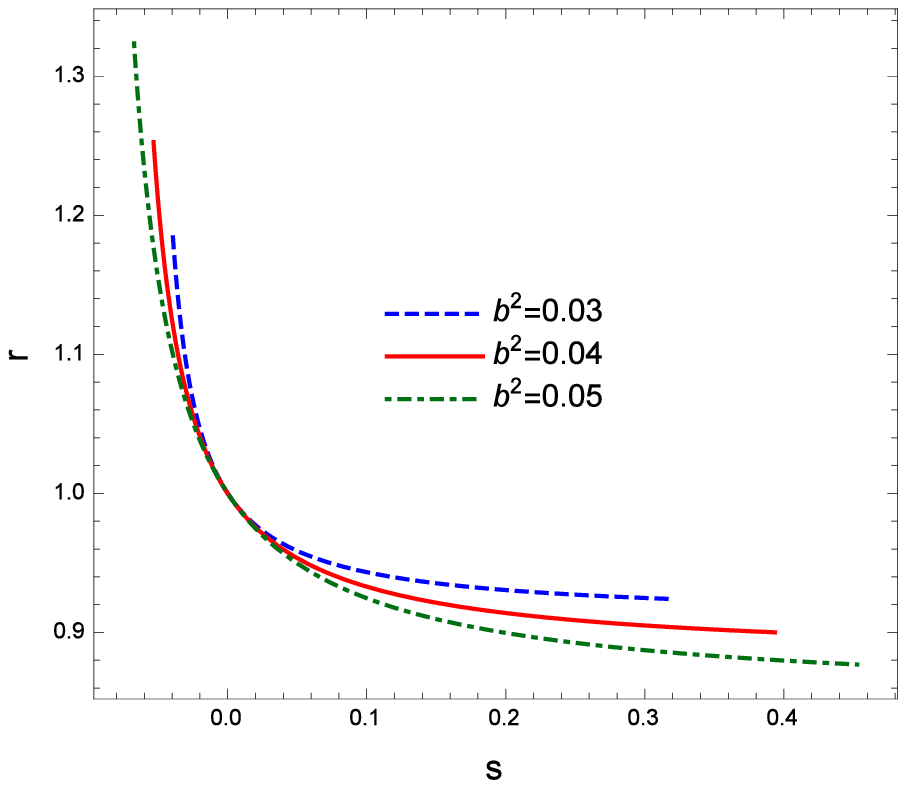}
\caption{The evolution of the statefinder parameter $r$ versus $s$ for sign-changeable interacting THDE with Ricci horizon as IR cutoff.}\label{rs-z5}
\end{center}
\end{figure}
\begin{figure}[htp]
\begin{center}
\includegraphics[width=8cm]{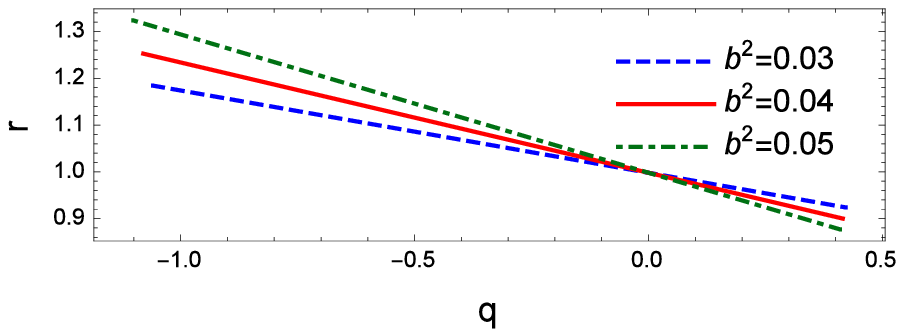}
\caption{The evolution of the statefinder parameter $r$ versus the deceleration parameter $q$ for sign-changeable interacting THDE with Ricci horizon as IR cutoff.}\label{rq-z5}
\end{center}
\end{figure}
\begin{figure}[htp]
\begin{center}
\includegraphics[width=6cm]{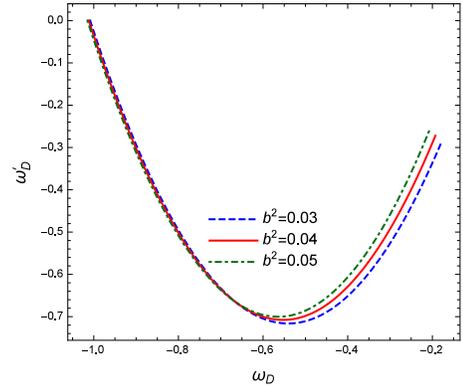}
\caption{The $\omega_D-{\omega}^{\prime}_{D}$ diagram for sign-changeable interacting THDE with Ricci horizon as IR cutoff.}\label{ww-z5}
\end{center}
\end{figure}
\section{Closing remarks}
A Bianchi type I universe was considered filled by DM and DE
interacting with each other throughout a sign-changeable
interaction. In our study, THDE plays the role of DE and various
IR cutoffs, including the Hubble, event and particle horizons as
well as the GO and Ricci cutoffs, have been used to study the
evolution of the universe. We tried to present a comprehensive
study by addressing diverse parameters such as $q$, $H$, $r$,
${\omega}^{\prime}_{D}$ and etc. Although suitable dynamics can be
obtained for the models, the classical stability analysis
($v_s^2$) shows that the models are always unstable at the
$z\rightarrow-1$ limit. It is also worthwhile mentioning that the
models may show stability by themselves at the classical level for
the current universe (where $z\rightarrow0$), depending on the
unknown parameters of model such as $\delta$.

\acknowledgments{We thank Shiraz University Research Council. This
work has been supported financially by Research Institute for
Astronomy \& Astrophysics of Maragha (RIAAM), Iran. In addition, the
work of KB was partially supported by the JSPS KAKENHI Grant Number
JP 25800136 and Competitive Research Funds for Fukushima University
Faculty (18RI009).}

\end{document}